# Optimal Salaries of Researchers with Motivational Emergence

Eldar Knar[1]

Tengrion, Astana, Republic of Kazakhstan
eldarknar@gmail.com
https://orcid.org/0000-0002-7490-8375

**Abstract**

In the context of scientific policy and science management, this study examines the system of nonuniform wage distribution for researchers. A nonlinear mathematical model of optimal remuneration for scientific workers has been developed, considering key and additive aspects of scientific activity: basic qualifications, research productivity, collaborative projects, skill enhancement, distinctions, and international collaborations. Unlike traditional linear schemes, the proposed approach is based on exponential and logarithmic dependencies, allowing for the consideration of saturation effects and preventing artificial wage growth due to mechanical increases in scientific productivity indicators.

The study includes detailed calculations of optimal, minimum, and maximum wages, demonstrating a fair distribution of remuneration on the basis of researcher productivity. A linear increase in publication activity or grant funding should not lead to uncontrolled salary growth, thus avoiding distortions in the motivational system.
The results of this study can be used to reform and modernize the wage system for researchers in Kazakhstan and other countries, as well as to optimize grant-based science funding mechanisms. The proposed methodology fosters scientific motivation, long-term productivity, and the internationalization of research while also promoting self-actualization and ultimately forming an adequate and authentic reward system for the research community.

Specifically, in resource-limited scientific systems, science policy should focus on the qualitative development of individual researchers rather than quantitative expansion (e.g., increasing the number of scientists). This can be achieved through the productive progress of their motivation and self-actualization.

**Keywords:** nonlinear salary model, researcher remuneration, grants, scientist motivation, self-actualization, economics of science, golden ratio.

**Declarations and Statements:**
No conflicts of interest
This work was not funded
No competing or financial interests
All the data used in the work are in the public domain.
Generative AI (LLM or other) was not used in writing the article.
Ethics committee approval is not needed (without human or animal participation).

---

[1] Fellow of the Royal Asiatic Society of Great Britain and Ireland

# 1. Introduction

In December 2024, the Ministry of Science and Higher Education of Kazakhstan issued a decree titled

*"On the Approval of the Rules for the Remuneration of Researchers in State Scientific Organizations and Public Organizations of Higher and (or) Postgraduate Education Performing Government Orders."*

Later, in February 2025, this decree was republished for public discussion in a slightly modified form[2].

We do not focus on the actual content of the decree but note that its release sparked widespread discussions and led to negative reactions from certain segments of Kazakhstan's scientific community[3]. In particular, the Council of Young Scientists of the National Academy of Sciences of the Republic of Kazakhstan directly addressed the issue on social media, expressing strong disagreement with the decree.

From the perspective of the critics, the implementation of Decree No. 592 will result in a sharp reduction in already modest salaries for researchers. This means that not only will incentive mechanisms disappear, but the very motivation to engage in scientific work may also be lost, especially for so-called "young scientists" (officially defined in Kazakhstan as those under 40).

The primary concern revolves around the revision and reassessment of bonuses and coefficients in the scientific wage system.

To clarify, in Kazakhstan, only a limited number of researchers and scientific personnel receive a guaranteed and stable salary. This includes administrative and managerial personnel, so-called "leading scientists," and researchers affiliated with certain "fundamental" institutes. Moreover, for the vast majority of researchers, research or commercialization grants (science-to-market) serve as their only form of scientific employment.

Therefore, within the grant system, salary levels are critically important. From the perspective of scientific policy, salary levels determine the outcomes of the "war for talent[4]." This issue is particularly relevant for Kazakhstan, as it grapples with the internal migration of its creative class and scientific talent, which tends to dissipate into the global scientific community.

---

[2] On Amendments to the Order of the Acting Minister of Science and Higher Education of the Republic of Kazakhstan dated December 25, 2024 No. 592 "On Approval of the Rules for Remuneration of Researchers of State Scientific Organizations and State Organizations of Higher and (or) Postgraduate Education Fulfilling State Orders"»

[3] Since we have not conducted a corresponding sociological survey, we cannot say whether this is a larger or smaller part of the scientific community

[4] Marini, G. (2024). Brexit and the War for Talents: Push & pull evidence about competitiveness. Higher Education. https://doi.org/10.1007/s10734-024-01186-1

We do not resort to banal economic statements suggesting that high wages automatically lead to high productivity—especially since this is not always the case.

However, we can assert with certainty that the labor of researchers should be compensated at an elite level-distinctly and unequivocally above the national average. This is because science does not merely produce innovations and discoveries; it creates something even more fundamental—competence and culture.

These are the strategic assets of the state and society, forming the foundation of all progress.

The newly issued Kazakh decree appears to be an attempt to reduce researchers' salaries, likely stemming from a misinterpretation of the state's new budgetary objectives, which are framed as "necessity for cost savings."

The issue of wage limitations is an age-old problem in governance and labor economics. In Adam Smith's economic models, the problem was resolved quite simply: pay as little as possible and make workers produce as much as possible.

However, this approach led to the emergence of Marxism and communism, where wages were low but work expectations were also relatively lax—except in cases of coercion, such as in the Siberian labor camps.

With respect to scientific labor, the minimum salary threshold for researchers represents a boundary condition between motivation and demotivation at its lowest limits. This signifies a phase transition between preventive demotivation and the initial stages of scientific motivation growth (excluding exceptional cases of self-funded or low-funded scientific enthusiasm).

However, the fundamental problem is that a developed economy and a progressive scientific system cannot be built through cost-cutting alone. This requires substantial investments (up to 3–5% of GDP), and most importantly, these high expenditures must be efficient and rational within the framework of continuous and well-structured scientific policy and management[5].

We argue that high expenditures on science and researchers are, without a doubt, efficient and rational investments. This is evident from the statistical data and correlation trends between scientific development and national progress.

Many people fail to see the tangible effects of science, and for this reason, they lack the vision and understanding of how these effects truly manifest.

In general, we do not attempt to prove the role of science in the state and society here[6]. Instead, we simply state a basic fact: without adequate and elite financial support for science and individual scientists, it is materially and immaterially impossible to create a "high-quality and progressive scientific system" and, even more so, a "research hub" in Kazakhstan.

---

[5] Eldar Knar (2024). Homeopathic Modernization and the Middle Science Trap: conceptual context of ergonomics, econometrics and logic of some national scientific case. arXiv preprint arXiv:2411.15996.

[6] Eldar Knar (2024). HOMO SCIENCE. Cambridge Open Engage. doi:10.33774/coe-2024-l2m9d This content is a preprint and has not been peer-reviewed.

However, it is important not to take this to extremes. When salaries become excessively high, motivational vectors shift. Beyond a certain threshold, salaries cease to play a constructive role in individual research productivity, instead becoming a resistant or even destructive factor.

At the core of scientific activity lies motivation. Motivation can be approximately defined as a specific form of human consciousness, interpreted through the drive for knowledge, exploration, recognition, social significance, and self-expression[7]. Motivation can take various forms, but it is fundamentally the foundation of scientific innovation and progress. However, motivation must also be materially supported.

Moreover, excessive financial or material support can, in certain cases, transform motivation into demotivation. For example, after receiving a Nobel Prize or another prestigious award, a researcher's scientific activity often declines sharply or moderately (this is not a strict rule but rather an observation).

Naturally, this raises the question of what constitutes an optimal, minimum, and maximum salary for a researcher—one that remains a constructive, organizing, and stimulating factor that always remains on the positive side of motivation. Here, the challenge of balancing intentionality and materiality in R&D emerges.

For the optimal salary, we refer to a level of remuneration that ensures emotional satisfaction. This is often described as "emotional salary[8]." However, we believe that emotional salary is not strictly defined by its monetary value alone.

High emotional satisfaction is achieved when a scientist navigates a certain "corridor of opportunities" to reach or exceed this level. In this context, not only does the final salary amount matter but also the range of pathways available for achieving it.

Psychological motivation and emotional satisfaction with one's salary—achieved through self-actualization—are universal concepts that are not limited to science alone.

In this sense, the career path of a researcher can be interpreted through the following hypothetical framework:

---

[7] Junça Silva, A., Burgette, A. R., & Fontes da Costa, J. (2024). Toward a Sustainable World: Affective Factors Explain How Emotional Salary Influences Different Performance Indicators. Sustainability (Switzerland), 16(5). https://doi.org/10.3390/su16052198

[8] Aviles-Peralta, Y. (2024). Emotional Salary: beyond traditional compensation. *Región Científica*. https://doi.org/10.58763/rc2024191

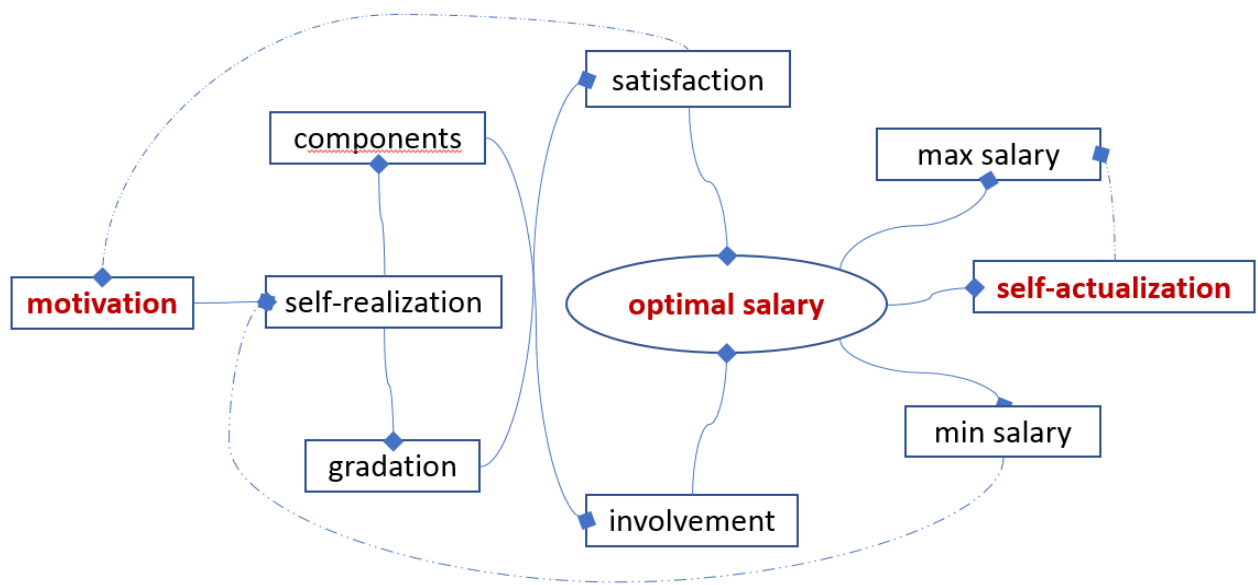

*Figure 1.* Motivation, Optimal Salary, and Self-Actualization**.**

Thus, motivation inevitably leads to the necessity of self-actualization. Scientific self-actualization is interpreted through achieving maximum success in various components (recognition, achievements, results, competencies, distinctions, collaboration, etc.) and within the grading system (scientific titles, degrees, positions, and ranks). The achieved milestones in these areas shape a certain level of satisfaction and engagement in scientific activity. The optimal levels of satisfaction and engagement are interpreted through the achieved salary level, which may be at an optimal level (both objectively and subjectively perceived) or fluctuate at the upper threshold limits.

Ideally, all of this leads to the state of a researcher's self-actualization[9].

Thus, having all researchers "stagnate" on a single universal base salary is the most effective way to destroy competition, motivation, and functionality in science[10]. Therefore, the more factors and components we account for in a researcher's salary, the greater their scientific motivation, functionality, and, ultimately, effectiveness and productivity.

A scientist may underperform in distinctions (insignia), but they have the opportunity to compensate for this local disadvantage through strengths in competencies or collaborations. Alternatively, a researcher who is an introvert[11] and is minimally involved in collaboration can compensate for this shortcoming through distinction. Overall, the diversity of components allows scientists to develop various winning strategies to maximize their salary levels.

---

[9] scientific self-actualization is identical to self-actualization in the notation of Maslow's pyramid of needs, but with some additions and variations

[10] Woolston, Chris. (2021). Stagnating salaries present hurdles to career satisfaction. Nature. 599. 519-521. 10.1038/d41586-021-03041-0

However, the most successful strategy for a researcher is, of course, performance and effectiveness. This is the dominant strategy, and success in this area fully compensates for all other deficiencies and shortcomings.

Thus, while salary has quantitative boundary conditions, from a qualitative and synthetic perspective, salary has almost unlimited potential for motivational growth.

These considerations form the basis of the proposed nonlinear mathematical model for determining the optimal, minimum, and maximum salaries for researchers.

**2. Literature Review**

The remuneration system for researchers is one of the key topics in science management, the knowledge economy, and research policy. The issue of fair and motivating compensation for scientific personnel is widely discussed in both academic research and national science support programs. This review examines the main approaches to evaluating researchers' salaries, including traditional linear models, modern nonlinear approaches, and principles of incentivizing scientific activity.

Historically, several models of researcher remuneration have been developed. Traditionally, salaries are based on a fixed rate that depends on qualification level, experience, and academic rank[12]. In several countries, a grading system is applied, in which researchers are classified into categories (assistant, associate professor, professor) with fixed salary scales[13].

However, as researchers have pointed out[14], the linear salary model does not stimulate productivity, as it does not take into account individual achievements and contributions to science. Moreover, it fails to reflect differences in researchers' effectiveness and does not incentivize international collaboration.

As a result, some countries have implemented hybrid systems that combine a base salary with performance-based bonuses. For instance, the U.S. uses a *tenure-track* system, where salaries increase based on the number of publications and acquired grants.

---

[11] Eldar Knar (2024). Recursive introversion, iterative extroversion and transitive ambiversion. arXiv preprint arXiv:2501.00043

[12] OECD (2018). Education at a Glance 2018: OECD Indicators. OECD Publishing.

[13] Marginson, S. (2017). The Worldwide Trend to High Participation Higher Education. Higher Education, 72(4), 413-434.

[14] Altbach, P. G., Reisberg, L., & Rumbley, L. E. (2019). Trends in Global Higher Education: Tracking an Academic Revolution. UNESCO Publishing.

In recent years, nonlinear salary models have gained traction, recognizing the complex dynamics of scientific activity. It has been proven that scientific productivity grows nonlinearly and exhibits saturation effects[15].

Thus, modern scientific economics requires models with nonlinear constraints[16].

Therefore, contemporary scientific economies demand hybrid and nonlinear remuneration models that combine fixed, performance-based, and collaborative components. Implementing constraints on grant-based payments and the **H-index** helps balance the motivational system.

**3. Results**

**3.1.** *Screening*

As a result of the conducted research, a nonlinear mathematical model of the optimal salary for a researcher was developed, taking into account key aspects of scientific activity. The derived model integrates six fundamental components:

*Fig 2.* Six dominant components of a scientist's salary

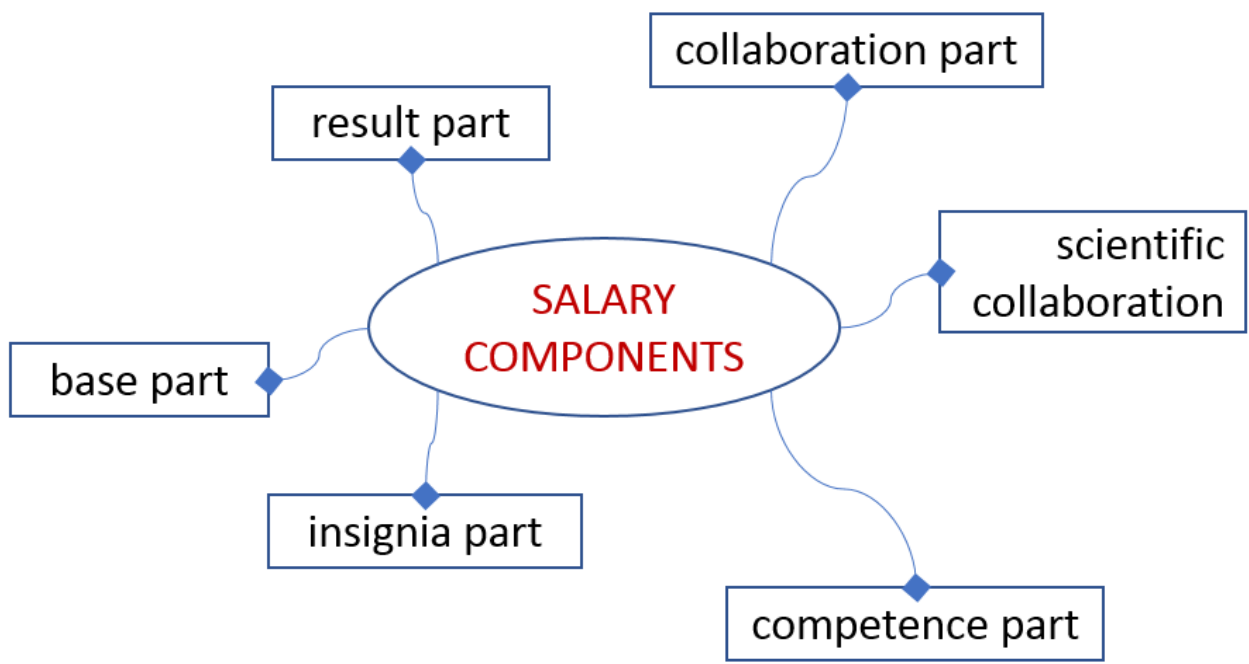

---

[15] Stephan, P. (2012). How Economics Shapes Science. Harvard University Press
[16] Franzoni, C., Scellato, G., & Stephan, P.E. (2011). Changing Incentives to Publish. Science, 333, 702 - 703.

Each of these components contributes to the final remuneration of a researcher, ensuring a fair and motivating salary system. The meaning and content of the components are interpreted through the following table:

*Table 1.* Functions, Attributes, and Criteria of Salary Components

| Function | Attribute | Criteria |
|---|---|---|
| $S_b$ | Base Salary Component | Depends on experience, qualifications, and position |
| $S_r$ | Performance-Based Component | Accounts for publications, citation index, and grant funding acquired |
| $S_c$ | Collaborative Component | Rewards participation in scientific teams and interdisciplinary projects |
| $S_s$ | Competency-Based Component | Reflects skill improvement, certifications, and professional experience |
| $S_i$ | Distinction-Based (Insignia) Component | Rewards honors, awards, prestigious memberships, and scientific recognition |
| $S_g$ | Scientific Collaboration Component | Encourages participation in international projects and integration into the global scientific community |

Thus, the resulting salary of a researcher can be expressed in the following additive form:

$$S = S_b + S_r + S_c + S_s + S_i + S_g$$

This is a simple yet fundamental superposition of salary components for a researcher.

Let us examine these components in terms of their specific contributions.

### 3.2. *Base Salary Component*

The base salary component should explicitly account for work experience, qualifications, and a guaranteed base rate.

Work experience T is interpreted as an increase in salary with increasing experience but with a saturation effect (i.e., without linear growth). The qualification level L represents increasing qualifications (master's, PhD, Doctor of Science), which should also increase salary but without excessive dominance. The guaranteed base salary $W_0$ serves as the minimum salary a researcher receives.

Thus, we need to construct a function $S_b(T,L)$ that ensures the following:

*Logical salary growth with experience but with a parallel saturation effect (interpreted via a logarithmic function),*

*An increase in work experience but moderate to prevent unchecked linear escalation,*

*The impact of the qualification level without disproportionate dominance,*

*Adjustability of salary growth through parameter tuning.*

To account for diminishing returns on work experience, the function is interpreted via logarithmic dependence:

$$f(T) = \ln\left(1 + \frac{T}{T_0}\right)$$

where

T is the number of years of work experience,

$T_0$ is a normalization constant defining the dependency's nature (e.g., $T_0$=10 means that significant changes occur in the first ten years of research, after which growth slows or restructures).

A simple logarithmic function exhibits weak nonlinearity. Therefore, we introduce an exponential coefficient β, which enhances the saturation effect:

$$f(T)^\beta = \left(\ln\left(1 + \frac{T}{T_0}\right)\right)^\beta$$

A condition β>1 ensures decelerated salary growth. For instance, at β=1.2, the salary increase is smoother than that at β=1.

To link this enhanced function with the base salary $W_0$, we introduce a correction factor α:

$$S_b = W_0 + \alpha \left(\ln\left(1 + \frac{T}{T_0}\right)\right)^\beta$$

where

α is an adjustment coefficient that regulates the impact of experience on salary.

Salary should also increase proportionally with qualification level L, but this increase should not be infinite. Introducing the corresponding coefficient results in:

$(1+\lambda L)$

where

L is the qualification level (e.g., 1 = Master's, 2 = PhD, 3 = Doctor of Science),

λ is a coefficient that determines the influence of the qualification level.

Thus, the final equation for the base salary component is as follows:

$$S_b = W_0 + α(1 + λL)(ln(1 + \frac{T}{T_0}))^β$$

The logarithmic dependency ensures rapid salary growth in the early years, followed by a gradual slowdown. The exponential coefficient β regulates the degree of nonlinearity. If β>1, growth is slower; if β<1, growth is faster (not used here, as it contradicts the idea of saturation). The qualification coefficient λ accounts for the effect of higher education without excessive influence.

Example Calculations,

Case 1: Master's degree, First Year of Work

$S_b$ = 190000·(1+0.05ln(1+0/5))1.2×(1+0.1×1)

$S_b$ = 190000×(1)1.2×1.1 = 209000 KZT

Here, 209,000 KZT represents the minimum base salary for a researcher.

Case 2: Doctor of Science, 40 Years of Experience

$S_b$ = 190000·(1+0.05ln(1+40/5))1.2×(1+0.1×3)

$S_b$ = 190000·(1+0.05×2.1972)1.2×1.3

$S_b$ = 190000×(1.10986)1.2×1.3

$S_b$ = 190000×1.1543×1.3=279911 KZT

Here, 279,911 KZT represents the maximum base salary for a researcher.

This formula accounts for experience without excessive linear growth, incorporates the qualification level proportionally without sharp jumps, provides a logical salary range from minimum to maximum values and allows adjustments via parameters α, β, and λ to fit different salary systems.

Fine-tuning will depend on further analysis and national science policy regarding research incentives.

### 3.3. Performance-Based Component

The performance-based component of a researcher's salary should account for key indicators of scientific productivity. Specifically, the number of indexed publications (P) is used to encourage active publishing and research productivity, and the H-index or K-index[17] reflects the quality and impact of indexed scientific publications and grant funding (G), which is interpreted through participation in research grants and increases the overall contribution of a scientist to the field.

When assessing the performance-based component, it is essential to consider the following effects:

Nonlinearity effect – a high number of scientific publications or high index values should not lead to an excessively high salary increase;

Diminishing returns – each subsequent scientific achievement should provide a decreasing level of salary growth.

Grant Limitations – Grants have significant weight, but the number of grants per researcher should be limited, with a parallel diminishing return on salary.

On the basis of these effects, the performance-based component can be expressed as a local superposition:

$$S_r = S_P + S_H + S_G$$

where
$S_P$ is the contribution of publication activity,
$S_H$ is the contribution of the H-index or K-index,
$S_G$ is the contribution of grant activity.

To account for the nonlinear impact of scientific publications, we introduce a power-law dependency:

$$S_P = \gamma_1 P^{\delta_1}$$

where
$\gamma_1$ is the coefficient regulating the contribution of publications to salary,
$\delta_1$ is the nonlinearity parameter ($1<\delta_1<1.2$), allowing the influence of publications to accelerate or decelerate. If $\delta_1>1$, publications have a greater impact at the early stages of a research career, but the effect saturates over time.

---

[17] Eldar Knar (2024). Recursive index for assessing value added of individual scientific publications. arXiv preprint arXiv:2404.04276.

The H-index (or K-index) measures the citation impact of publications. We use a similar power function:

$$S_H = \gamma_2 H^{\delta_2}$$

where
$\gamma_2$ is the coefficient influencing the H- or K-index contribution,
$\delta_2 \approx 1$ allows for slightly slowed growth.

Grants are among the most complex components, as they should not dominate the salary structure. In many scientific systems, grants have become the primary source of research funding for the majority of researchers.

To model grants, we apply the following principles:
Limited impact of grant quantity - modeled using the golden ratio ($\phi=1.618$),
Normalization of Grant Amounts - expressed in the national currency,
The corrective coefficient (gif - grant impact factor) limits the excessive influence of grants.

Thus, the formula for the grant contribution is given by:

$$S_G = \gamma_3 \left( \phi^{\frac{Gp}{\max Gp}} \cdot \left( \frac{G}{10^6} \right)^{gif} \right)$$

where
$G_p$ is the number of grant projects (limited to 3),
G is the total amount of attracted grants (in local currency),
$\gamma_3$ is the grant weight coefficient,
$\phi=1.618$\phi = 1.618$\phi=1.618$ is the golden ratio,
gif (grant impact factor) is a corrective parameter limiting grant influence (0.7–0.8).

Final Formula for the Performance-Based Component

$$S_r = \gamma_1 P^{\delta_1} + \gamma_2 H^{\delta_2} + \gamma_3 \left( \phi^{\frac{Gp}{\max Gp}} \cdot \left( \frac{G}{10^6} \right)^{gif} \right)$$

Extreme cases for researchers:

Case 1: Master's Graduate with No Research Experience
If a researcher has no publications, grants, or citation indices, their performance-based salary component is as follows:

$S_r = 15000×0^{1.05} + 10000×0^{1.1} + 20000×0 = 0$

Case 2: Senior Researcher (Doctorate) with Maximum Achievements
A researcher with 100 indexed publications, an H-index of 50, and three grants of 50 million in local currency receives the following:

$S_P = 15000×100^{1.05} = 15000×125.89 = 1\ 888\ 350$

$S_H = 10000×50^{1.1} = 10000×63.10 = 631\ 000$

$S_G = 20000×(1.618×9.42) = 20000×15.24 = 304\ 800$

Thus, the total performance-based component for a senior researcher is as follows:

$S_r = 1888350 + 631000 + 304800 = 2\ 824\ 150$ KZT

The proposed formula

$$S_r = \gamma_1 P^{\delta_1} + \gamma_2 H^{\delta_2} + \gamma_3 \left( \phi^{\frac{Gp}{\max Gp}} \cdot \left( \frac{G}{10^6} \right)^{gif} \right)$$

ensures nonlinear salary growth as publications and citation indices increase while incorporating a saturation effect. Grants contribute in a controlled manner, governed by the golden ratio and a power function, balancing publication activity, citation impact, and research funding. This approach guarantees fair and authentic compensation for scientific productivity.

The collaborative, competency-based, insignia-based, and research-collaborative components are structurally and logically identical to the basic and performance-based components.

Therefore, we limit our analysis to the final mathematical representation and its numerical interpretation.

### 3.4. Collaborative Component

The final equation for the collaborative component is as follows:

$$S_c = \lambda_1 (1 - e^{-\mu_1 C})$$

where

C is the number of internal research projects within a scientific organization or an accredited entity,
$\lambda_1$=50,000 (assumed value),
$\mu_1$=0.1 (assumed value).

Salary threshold range:

Minimum: $S_c = 0$,
Maximum: $S_c = 43,233$ KZT at C = 20 (assumed value).

### 3.5. Competency-based Component

The final equation for the competency-based component is as follows:

$$S_s = \lambda_2 \left(1 - e^{-\mu_2 K}\right)$$

where
K is the number of professional development courses or certifications obtained by a researcher,
$\lambda_2$=40,000 (assumed value),
$\mu_2$=0.15 (assumed value).

Salary threshold range:

Minimum: $S_s$=0,
Maximum: $S_s$=31,075 KZT at K = 10 (assumed value).

### 3.6. Insignia-Based Component

The final equation for the insignia-based component:

$$S_i = \lambda_3 \left(1 - e^{-\mu_3 I}\right)$$

where
I represents the number and quality of awards, memberships, and honors,
$\lambda_3$=70,000 (assumed value),
$\mu_3$=0.1 (assumed value).

Salary threshold range:

Minimum: Si=0,
Maximum: Si=44,248 KZT at I=10 (assumed value).

**3.7. Scientific Collaboration Component**

The final equation for the scientific collaboration component is as follows:

$$S_g = \lambda_4 (1 - e^{-\mu_4 SC})$$

where:
SC is the number of international research projects,
$\lambda_4$=100,000 (assumed value),
$\mu_4$=0.2 (assumed value).

Salary threshold range:

Minimum: $S_g$=0,
Maximum: $S_g$=86,467 KZT at SC=10 (assumed value).

**3.8. Optimization and Extremum**

The minimum salary of a researcher is as follows:

$S_{min}$ = 209000+0+0+0+0+0 = **209 000** KZT

The maximum salary, considering all parameters at their upper limits:

$S_{max}$= 279911+3128133+43233+31075+44248+86467 = **3 613 066** KZT

To estimate the optimal salary for a researcher, we apply a logarithmic optimization criterion. The logarithmic mean accounts for relative changes, providing a balanced growth perspective and reducing sensitivity to extreme values.
Using this approach, the optimal salary level for a researcher is estimated as follows:

$S_{opt}$ ≈ **868 982** KZT

Here, the optimal salary reflects an emotionally satisfying income level, often referred to as the "emotional salary."
The optimal, minimum, and maximum salaries of researchers with motivational emergence can be represented in a graphical format:

*Figure 3.* Overview of Salary Parameters for Researchers

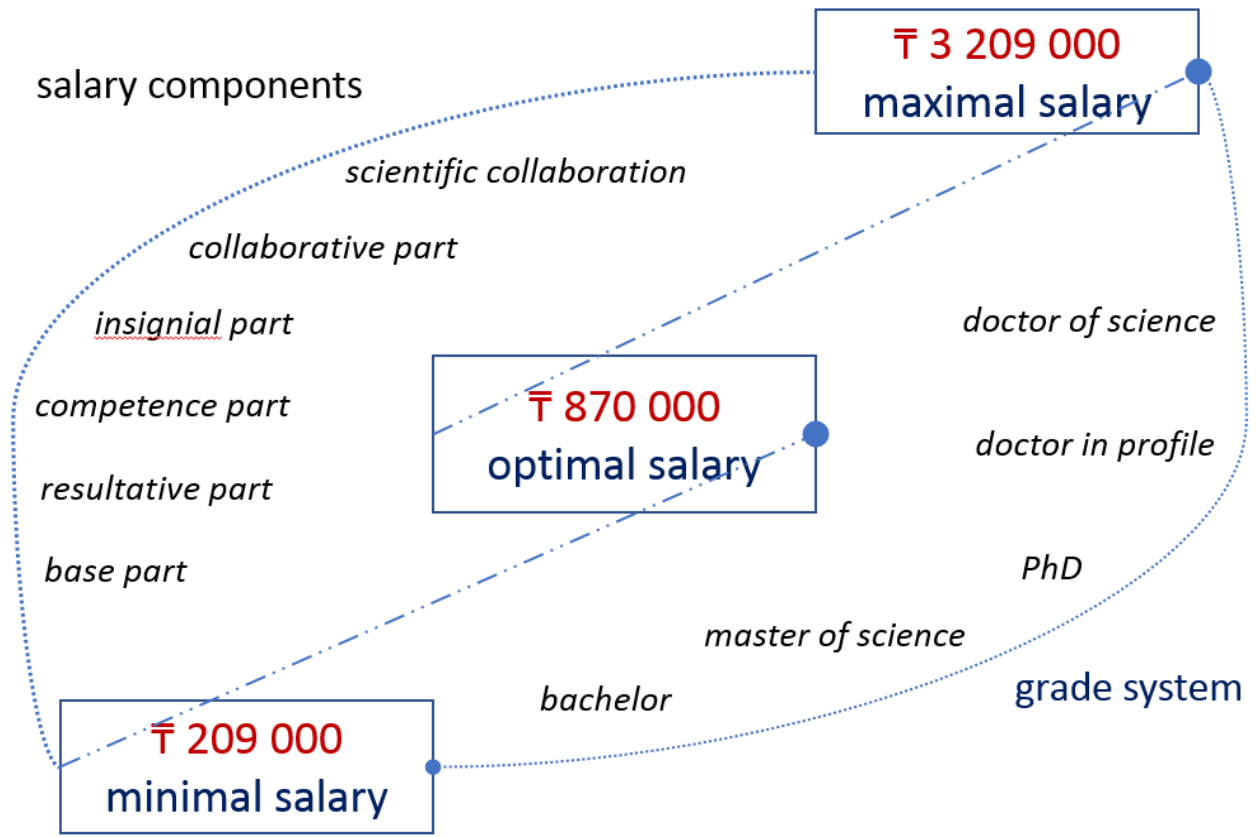

This visual representation highlights that both the optimal and maximum salary levels align well with Kazakhstan's financial capabilities, considering its increasing science budget and resource limitations, including human capital.

**4. Discussion**

Scientific activity is highly specific and fundamentally different from traditional labor markets for several reasons. Scientific work operates on a 24/7 basis, often continuing beyond official working hours or even extending into subconscious processing during sleep.

Scientific labor possesses cognitive value—it is a form of compensation for intelligence and creativity, both of which significantly contribute to societal competence and culture. However, research productivity and efficiency are only possible at a very high level of motivation. Thus, achieving a motivational balance in science through fair compensation is essential for the progress of both science and society.

Scientific work is a continuous and permanent endeavor. Like administrative officials, teachers, or doctors, researchers require stable, not situational, compensation as part of their base salary.

Moreover, scientific labor is additive—it accumulates over time. As such, a researcher's total salary should follow the principle of additivity, integrating multiple components that reflect different aspects of their contributions.

Thus, the remuneration of researchers must be authentic and adequate, particularly in the context of incentives, productivity, effectiveness, and professional development.

Science does not always yield immediate or short-term results. Therefore, the salary of a researcher should always be interpreted as an investment in the future—just as one would save for a child's education. Failing to do either ultimately means forfeiting any potential future development.

In summary, the methodology and mechanisms for determining researchers' salaries must be unique rather than universal or standardized.

In most traditional professions, salary is a direct payment for labor already performed. However, in scientific activity, salary includes an additional component: an investment in future research outcomes.

This investment creates conditions for the full realization of a researcher's intellectual and creative potential. It requires a balance between material incentives and an environment conducive to intellectual and creative self-actualization. Of course, investment-based salary models exist in other fields, but in science, this factor is expressed most clearly.

In other words, a researcher's salary is not only a reward for current research efforts but also a preventive payment for the extra effort required to achieve new scientific results. However, such extra efforts are only possible and sustainable under conditions of genuine and constructive internal motivation.

According to self-determination theory[18] (SDT), internal motivation makes a researcher maximally productive.

Therefore, salary levels must be set high enough to maintain motivation at a sufficiently elevated level, but not excessively high, as this could lead to motivational inversion (a phenomenon where excessive compensation results in decreased intrinsic motivation).

Motivation is neither an abstract nor an ephemeral concept—it can, at least to some extent, be quantified and evaluated.

According to expectancy theory[19], the strength of motivation can be interpreted through the following formula:

---

[18] Deci E.L., Ryan R.M. (2008) *Self-determination theory: A macrotheory of human motivation, development and health.* Canadian Psychology, V 49, p 182-185.

[19] Victor H. Vroom (1994) Work and Motivation 1st Edition. Jossey-Bass, p 432, ISBN-10 0787900303

$$\textit{Motivational Force (MF)} = \textit{Expectancy (E)} \times \textit{Instrumentality (I)} \times \textit{valence (V)}$$

where:

MF - motivational force of an individual (in this case, a researcher),

E - expectancy, i.e., the perceived probability that effort will lead to a meaningful scientific outcome,

I - instrumentality, i.e., the degree to which achieving a result is essential for obtaining material or nonmaterial rewards;

V - valence, i.e., the perceived value or significance of the expected outcome.

For example, if a highly motivated researcher is fairly confident that their research will lead to a breakthrough discovery (E = 0.8), expects that their discovery will be properly recognized (I = 0.9), and believes that achieving this recognition will lead to career advancement and salary growth (V = 0.9), then their motivational force is as follows:

$$MF = 0.8 \times 0.9 \times 0.9 = 0.65$$

This is considered a sufficiently high level of motivation.

We have determined the optimal salary level for a researcher to be 868,982 KZT. However, it is essential that this optimal salary is correlated with motivational force (MF).

We propose that the optimal salary should correspond to or exceed the motivation force threshold at approximately MF = 0.5.

Thus, the optimal salary should always satisfy the following condition:

$$0.5 < MF < 1$$

Within this range, in accordance with Herzberg's two-factor motivation theory[20], both hygiene factors (salary, job security) and motivating factors (achievements, recognition, professional growth) reach their optimal balance.

Within this range, the optimal research salary becomes emergent—it is no longer just a simple sum of salary components but rather a significant activator of nonmaterial and intrinsic research motivations.

---

[20] Frederick Herzberg, Bernard Mausner, Barbara Bloch Snyderman (2011) The Motivation to Work. Organization and Business. Transaction Publishers, p 180, ISBN 1412815541

The minimum and maximum salaries we calculated should be seen as necessary salary structures for researchers with varying motivations and self-actualization levels.

For an early-career master's degree, the dominant motivating factor is simply entry into the scientific field—that is, engagement in research and the opportunity to pursue a career in science. At this minimal level of motivation, a young researcher is generally content with a lower salary, with the expectation that it will increase as they progress along the academic career path.

For experienced and accomplished researchers, the maximum salary level serves as a form of recognition for their past scientific achievements. Here, the motivation is different—to continue research and transfer scientific knowledge to the next generation.

Moreover, for young scientists, the financial well-being of senior researchers with high salaries serves as an additional stimulus and motivator to pursue scientific careers.

Thus, an effective scientific salary system must meet the following key conditions:

Unique and specialized approach—scientific salaries should not be universal or standardized, as in other professions;

Recognition of future potential—scientific salaries should not only reward current work but also fund future discoveries;

Intrinsic motivation support—a researcher's internal motivation is the key factor in ensuring scientific progress;

The optimal balance between material and nonmaterial incentives—a salary must be high enough to drive motivation but not excessive—could lead to motivation inversion.

Dynamic salary components—salaries must consider multiple factors, such as experience, qualifications, performance, collaboration, distinctions, and international engagement;

Correlation with motivation levels—the optimal salary should match or exceed a motivational force of 0.5 to ensure sustainable productivity.

Thus, by properly structuring scientific salaries, we can create a highly productive and motivating environment for researchers that supports long-term progress in science and innovation.

5. **Conclusion**

In this study, a nonlinear mathematical model for the optimal salary of a researcher was developed, accounting for key aspects of scientific activity: basic qualifications, research productivity, collaborative projects, skill enhancement, awards, and international cooperation. Unlike traditional linear salary schemes, the proposed model relies on nonlinear dependencies with saturation effects, ensuring fair compensation distribution while stimulating not only quantitative but also qualitative indicators of scientific activity.

The final salary formula for a researcher is expressed as the sum of six key components:

$$S = S_b + S_r + S_c + S_s + S_i + S_g$$

where:

$S_b$ - base salary, depending on experience and qualification level,
$S_r$ - research productivity, including publications, h-index, and grants,
$S_c$ - participation in scientific collaboration within the organization,
$S_s$ - competencies acquired through certification and professional development,
$S_i$ - awards and honorary titles confirming scientific reputation,
$S_g$ - international cooperation and integration into the global research community.

The use of logarithmic and exponential saturation functions allows the model to account for diminishing returns and avoid imbalances in salary structure.

The calculations demonstrated a realistic salary range on the basis of research activity:

Minimum salary (young researcher with no significant achievements) – 209,000 KZT.
Maximum salary (leading researchers with high productivity) – 3,613,066 KZT.

A crucial principle in science is that base salaries should be lower than those of performance-based components. If the guaranteed base salary is too high, it can lead to complacency and scientific stagnation due to emotional and motivational saturation. However, this principle is already being maintained within Kazakhstan's scientific framework.

Unlike existing systems, which often increase salaries linearly with increasing metrics, the proposed model limits the influence of grant projects, publication counts, and awards, preventing artificial inflation of research indicators.

The proposed model represents a modern approach to structuring researcher salaries, combining mathematical rigor, economic efficiency, and social fairness.

The use of nonlinear dependencies allows the creation of a system that motivates not only the quantity but also the quality of scientific outcomes.

A balanced distribution of factors guarantees sustainable growth in research productivity without artificial distortion of metrics.

The application of nonlinear models enables the following:

*Increased motivation of researchers through well-designed incentives.*

*Limiting artificial inflation of publication numbers. Strengthening the internationalization of science through collaboration bonuses.*

The model is ready for practical implementation and could serve as the foundation for salary system reforms for researchers in Kazakhstan and other countries.

Thus, the proposed methodology not only enhances researcher motivation but also establishes a sustainable system for long-term scientific development. The implementation of this model can directly improve the global competitiveness of Kazakhstan's research sector.

Finally, we believe that in resource-limited scientific systems, science policy should prioritize the qualitative development of individual researchers over quantitative expansion (e.g., increasing the number of scientists). This can be achieved by enhancing researcher motivation and self-actualization to drive long-term productivity and progress.